\documentclass[11pt,a4paper]{article}

\advance\textheight by \topskip
\usepackage{thumbpdf,lmodern}
\usepackage{natbib}
\usepackage{eqnarray,amsmath}
\usepackage{amssymb}
\usepackage{url}
\usepackage[linesnumbered]{algorithm2e}
\usepackage{Sweave}

\providecommand{\keywords}[1]{\small\textbf{\textit{Keywords:}} #1}

\let\proglang=\textsf
\newcommand{\pkg}[1]{{\fontseries{b}\selectfont #1}}
\let\code=\texttt

\newcommand{\vect}[1]{\boldsymbol{#1}}

\newcommand{\diff}{\mathop{}\!\mathrm{d}}

\author{Ziwen An\\Queensland University of Technology\\ACEMS\thanks{Australian Research Council Centre of Excellence for Mathematical and Statistics Frontiers}
\and Leah F South\\Lancaster University
\and Christopher Drovandi\\Queensland University of Technology\\ACEMS}

\title{\pkg{BSL}: An \proglang{R} Package for Efficient Parameter Estimation for Simulation-Based Models via Bayesian Synthetic Likelihood}

\date{\today}

\begin{document}

\maketitle

\begin{abstract}
Bayesian synthetic likelihood (BSL) \citep{Price2018} is a popular method for estimating the parameter posterior distribution for complex statistical models and stochastic processes that possess a computationally intractable likelihood function.   Instead of evaluating the likelihood, BSL approximates the likelihood of a judiciously chosen summary statistic of the data via model simulation and density estimation.  Compared to alternative methods such as approximate Bayesian computation (ABC), BSL requires little tuning and requires less model simulations than ABC when the chosen summary statistic is high-dimensional.  The original synthetic likelihood relies on a multivariate normal approximation of the intractable likelihood, where the mean and covariance are estimated by simulation.  An extension of BSL considers replacing the sample covariance with a penalised covariance estimator to reduce the number of required model simulations.  Further, a semi-parametric approach has been developed to relax the normality assumption.  In this paper, we present an \proglang{R} package called \pkg{BSL} that amalgamates the aforementioned methods and more into a single, easy-to-use and coherent piece of software.  The \proglang{R} package also includes several examples to illustrate how to use the package and demonstrate the utility of the methods.
\end{abstract}


\section[Introduction]{Introduction} \label{sec:intro}

In the Bayesian framework, inference on the parameter $\vect{\theta} \in \vect{\Theta} \subseteq \mathbb{R}^{p}$ of a statistical model is carried out using the posterior distribution $p(\vect{\theta} | \vect{y})$, where $\vect{y}$ is the observed data. Bayes' theorem shows that by observing $\vect{y}$ through the likelihood function $p(\vect{y} | \vect{\theta})$, the prior knowledge $p(\vect{\theta})$ can be updated to provide the posterior distribution,

\begin{equation*}
p(\vect{\theta} | \vect{y}) = \dfrac{p(\vect{y} | \vect{\theta}) p(\vect{\theta})}{\int_{\vect{\Theta}} \,p(\vect{y} | \vect{\theta}) p(\vect{\theta}) \,\diff \vect{\theta}} \propto p(\vect{y} | \vect{\theta}) p(\vect{\theta}).
\end{equation*}

In most applications, the evidence $\int_{\vect{\Theta}} p(\vect{y} | \vect{\theta}) p(\vect{\theta})\,\diff \vect{\theta}$ involves high dimensional integration and is intractable. Recovery of the posterior distribution often relies on sampling methods, such as Markov chain Monte Carlo (MCMC) and sequential Monte Carlo (SMC) \citep{Del2006}.

However, in complex models, the likelihood function can be intractable or very expensive to evaluate. The terminology ``likelihood-free inference'' typically refers to inference techniques that do not require direct evaluation of the likelihood function, but rely on model simulations to approximate the likelihood in some way. One successful method is approximate Bayesian computation (ABC) \citep{Sisson2018}. ABC essentially estimates the intractable likelihood non-parametrically at $\vect{\theta}$ with a simulation $\vect{x} \sim p(\cdot | \vect{\theta})$. The raw dataset is usually reduced down to summaries with a carefully chosen summary statistic function $\vect{S}(\cdot) : \mathbb{R}^{\delta} \mapsto \mathbb{R}^{d}$, where $\delta$ and $d$ are the dimension of the raw data and summary statistics, respectively. Denote the observed and simulated summary statistics as $\vect{s}_{\vect{y}} = \vect{S}(\vect{y})$ and $\vect{s}_{\vect{x}} = \vect{S}(\vect{x})$, respectively. The ABC likelihood function is given by

\begin{equation*}
p_{\epsilon} (\vect{s}_{\vect{y}} | \vect{\theta}) = \int_{\mathcal{Y}} \,K_{\epsilon} (\rho(\vect{s}_{\vect{y}}, \vect{s}_{\vect{x}})) p(\vect{x} | \vect{\theta}) \,\diff \vect{x},
\end{equation*}

where $\rho(\cdot)$ is a discrepancy function, which measures the distance between the observed and simulated summary statistics under a certain metric, e.g.\ the Euclidean distance. $K_{\epsilon}(\cdot)$ is a kernel weighting function, usually an indicator function $I(\cdot < \epsilon)$ for convenience, which links distance and tolerance $\epsilon$. Here $\epsilon$ is used to trade-off between the bias and variance of the likelihood estimator, i.e.\ as $\epsilon$ approaches zero the bias reduces but the variance increases. In sampling methods like MCMC, a likelihood estimator with large variance can cause the Markov chain to get stuck and reduce efficiency \citep{Doucet2015}.

Due to the non-parametric nature of the ABC likelihood estimate, ABC can be very inefficient when the summary statistic is high dimensional \citep{Blum2010}, which is often referred to as the curse of dimensionality. \citet{Wood2010} proposes to use a multivariate normal distribution to approximate the likelihood function. Such an approximation is called the synthetic likelihood (SL). We later extend the term to not only the multivariate normal distribution but also other reasonable parametric approximations of the likelihood (\citet{Drovandi2015} provide a general framework for such methods). \citet{Price2018} analyse SL in the Bayesian framework and name it Bayesian synthetic likelihood (BSL).  MCMC is used to explore the parameter space. The paper uses extensive empirical results to show that BSL can outperform ABC even when the model summary statistics show small departures from normality. BSL not only scales better with the summary statistic dimension, but also requires less tuning and can be accelerated with parallel computing.  Recently, asymptotic properties of BSL have been derived under various assumptions.  \citet{Frazier2019} develop a result for posterior concentration and \citet{Nott2019} show that the BSL posterior mean is consistent and asymptotically normally distributed.

In the standard BSL approach, the most obvious computational drawback is the need to generate a large number of simulations $n$ for estimating a high-dimensional covariance matrix in the SL. Several strategies have been devised to reduce the number of simulations $n$ for estimating the synthetic likelihood. \citet{An2019} propose to use the graphical lasso to estimate a penalised covariance and provide an algorithm for selecting the penalty to ensure reasonable mixing when placed inside an MCMC algorithm. \citet{Ong2018a} consider the shrinkage estimator of \citet{Warton2008} in the context of variational Bayes synthetic likelihood to reduce the number of simualtions. \citet{Everitt2017} considers a bootstrap approximation to lower the variance of the SL estimator. Furthermore, the normality assumption of SL has also been put under inspection by \citet{An2018}, which considers a more flexible approximation of the synthetic likelihood using a semi-parametric estimator with a Gaussian copula.

There are existing packages in \proglang{R} for likelihood-free inference. For example, the \proglang{R} packages \pkg{abc} \citep{Rpkg:abc} and \pkg{EasyABC} \citep{Rpkg:EasyABC} exist for various ABC methods. The \proglang{R} package \pkg{synlik} \citep{Rpkg:synlik} implements the classic SL method of \citet{Wood2010}. The package provides diagonostic tools for the normal synthetic likelihood and incorporates MCMC to find the approximate posterior distribution. However, \pkg{synlik} is limited to only the standard SL approach. Outside of \proglang{R}, the \pkg{ABCpy} \citep{Dutta2017} package in \proglang{Python} includes most of the popular ABC algorithms and the standard SL approach.

Given the wide applicability of BSL \citep[e.g.][]{Karabatsos2018, Barbu2018}, it is important that BSL methods are directly accessible to practitioners.  In this paper we introduce our \pkg{BSL} \proglang{R} package, which implements the Bayesian version of synthetic likelihood and many of the extensions listed earlier together with additional functionality detailed later in the paper. The package is flexible in terms of the prior specification, the implementation of the model simulation function and the choice of the summary statistic function. Further, it includes several built-in examples to help the user learn how to use the package.

The rest of this paper is structured as follows. Section \ref{sec:methods} provides a brief descriptions of the statistical background for SL and two shrinkage approaches that are implemented in our package. In Section \ref{sec:package}, we introduce the main functionalities in \pkg{BSL} with an illustrative example. Section \ref{sec:summary} concludes the paper with a summary and further discussion.

\section{Bayesian Synthetic Likelihood} \label{sec:methods}

Following the notation in Section \ref{sec:intro}, we focus on reviewing three SL estimators to $p(\vect{s}_{\vect{y}} | \vect{\theta})$ and two shrinkage covariance estimation methods in Section \ref{subsec:bsl} to \ref{subsec:SL_shrinkage}. Then we briefly introduce other implementation details in Section \ref{subsec:others}.

The SL estimator can be viewed as an auxiliary likelihood function $p_{A} (\vect{s}_{\vect{y}} | \vect{\phi})$ where the subscript ``A'' denotes that we are using a parametric density approximation and $\phi$ is the parameter of this parametric family of densities.  To link this auxiliary likelihood to the actual likelihood, there is actually a functional relationship between $\vect{\phi}$ and $\vect{\theta}$, which may be denoted as $\vect{\phi} (\vect{\theta})$. However, we drop the dependence on $\vect{\theta}$ for notational convenience. The SL posterior is given by

\begin{equation*}
p_{A}(\vect{\theta} | \vect{s}_{\vect{y}}) \propto p_{A}(\vect{s}_{\vect{y}} | \vect{\phi}) p(\vect{\theta}).
\end{equation*}

Unfortunately, the mapping from $\vect{\theta}$ to $\vect{\phi}$ is typically unknown. However $\vect{\phi}$ can be estimated with simulations. The estimated synthetic likelihood is placed within an MCMC algorithm to sample from the corresponding approximate posterior of $\vect{\theta}$. Below we describe the synthetic likelihood estimators supported by our package, which amount to choosing the form of $p_{A}$ and the type of estimator of $\vect{\phi}$.

\subsection{Standard BSL Likelihood Estimator} \label{subsec:bsl}

Assume we have obtained a collection of $n$ simulations from the model at a proposed parameter value of $\vect{\theta}$, i.e.\ $\vect{x}_{1},\,\dots,\,\vect{x}_{n} \overset{iid}{\sim} p(\cdot | \vect{\theta})$. The corresponding summary statistics are denoted as $\vect{s}_{i} = \vect{S}(\vect{x}_i) \in \mathcal{S} \subseteq \mathbb{R}^{d}$ for $i=1,\,\dots,\,n$. Following \citet{Wood2010}, the SL $p(\vect{s}_{\vect{y}} | \vect{\theta})$ is assumed to be roughly multivariate normal. The classic SL estimator can be written as

\begin{equation} \label{eq:likelihood_BSL}
p_{sl} (\vect{s}_{\vect{y}} | \vect{\phi}_{sl}) = \mathcal{N}(\vect{s}_{\vect{y}} | \vect{\phi}_{sl}),
\end{equation}

where $\vect{\phi}_{sl} = (\vect{\mu}(\vect{\theta}), \vect{\Sigma}(\vect{\theta}))$ is estimated with sample mean and covariance $\hat{\vect{\phi}}_{sl} = (\vect{\mu}_n(\vect{\theta}),\, \vect{\Sigma}_n(\vect{\theta}))$

\begin{equation} \label{eq:mu_Sigma_n}
\begin{aligned}
\vect{\mu}_n(\vect{\theta}) = & \dfrac{1}{n} \sum_{i=1}^{n} \vect{s}_i \\
\vect{\Sigma}_n(\vect{\theta}) = & \dfrac{1}{n-1} \sum_{i=1}^{n} (\vect{s}_i - \vect{\mu}_n(\vect{\theta})) (\vect{s}_i - \vect{\mu}_n(\vect{\theta}))^{\top}.
\end{aligned}
\end{equation}

Although $\hat{\vect{\phi}}_{sl}$ is an unbiased estimator of $\vect{\phi}_{sl}$, $\mathcal{N}(\vect{s}_{\vect{y}} | \hat{\vect{\phi}}_{sl})$ is not an unbiased estimator of $\mathcal{N}(\vect{s}_{\vect{y}} | \vect{\phi}_{sl})$. However, empirical results demonstrate that the approximate posterior exhibits very weak dependence on $n$ \citep{Price2018}. Nevertheless, if $n$ is prohibitively small, then there can be significant Monte Carlo error in the MCMC approximation of the BSL target.

\subsection{Unbiased BSL Likelihood Estimator} \label{subsec:ubsl}

\citet{Ghurye1969} provide an unbiased estimator for the multivariate normal density based on independent simulations from it. \citet{Price2018} show the viability of this estimator by using it in place of the standard SL estimator within an MCMC algorithm. The estimator, denoted as uBSL, requires $n > d+3$ for unbiasedness to hold. The estimated auxiliary parameter $\hat{\vect{\phi}}_{go}$ can be written as $(\vect{\mu}_{n} (\vect{\theta}),\, \vect{M}_{n} (\vect{\theta}))$, where $\vect{M}_{n} (\vect{\theta}) = (n-1) \vect{\Sigma}_{n} (\vect{\theta})$. Definitions of $\vect{\mu}_n(\vect{\theta})$ and $\vect{\Sigma}_n(\vect{\theta})$ can be found in equation \eqref{eq:mu_Sigma_n}. The unbiased likelihood estimator is given by

\begin{equation} \label{eq:likelihood_uBSL}
\begin{aligned}
p_{go} (\vect{s}_{\vect{y}} | \vect{\phi}_{go}) = & (2\pi) ^{-d/2} \dfrac{c(d,\,n - 2)}{c(d,\,n - 1) (1 - 1/n)^{d/2}} |\vect{M}_{\vect{n}} (\vect{\theta})| ^ {-(n-d-2)/2} \\
& \Psi \Big(\vect{M}_{\vect{n}} (\vect{\theta}) - \dfrac{(\vect{s}_{\vect{y}} - \vect{\mu}_{n}(\vect{\theta})) (\vect{s}_{\vect{y}} - \vect{\mu}_{n}(\vect{\theta})) ^ {\top}} {1-1/n} \Big) ^ {(n-d-3)/2},
\end{aligned}
\end{equation}

where 

\begin{equation*}
c(k,v) = \dfrac{2 ^ {-kv/2} \pi ^ {-k(k-1)/4}} {\prod_{i=1}^{k} \Gamma\Big(\dfrac{v-i+1}{2}\Big)}
\end{equation*}

and

\begin{equation*}
\Psi(\vect{A}) =
\begin{cases}
|\vect{A}|& \text{if $\vect{A} > 0$}\\
0 & \text{otherwise}
\end{cases}.
\end{equation*}

In spite of the fact that the normality assumption is rarely true in practice, the approximate posterior distributions obtained by uBSL show less dependence on $n$ when compared to BSL \citep{Price2018}.

\subsection{A Semi-Parametric BSL Likelihood Estimator} \label{subsec:semibsl}

Although the estimators of equations \eqref{eq:likelihood_BSL} and \eqref{eq:likelihood_uBSL} are straightforward to compute, the normality assumption of the summary statistic could be restrictive in some cases. It is thus of interest to consider a more robust likelihood estimator that can handle non-normal summary statistics whilst not sacrificing much on computational efficiency.

Here we briefly describe the semi-parametric likelihood estimator of \citet{An2018}, which is aliased as ``semiBSL'' by the authors. The paper shows examples where BSL fails to produce accurate posterior approximations and provides additional robustness to non-normality of the summary statistics. The semiBSL estimator harnesses kernel density estimation (KDE) to non-parametrically estimate the univariate distributions of the summary statistics and uses the Gaussian copula to model the dependence amongst summaries.  This semi-parametric estimator provides additional robustness over the multivariate normal assumption and is quick to apply.

The estimator involves two aspects; modelling the marginals and modelling the dependence structure. We first describe the non-parametric aspect of KDE. Denoting the marginals of $\vect{s}_{\vect{i}}$ as $s_{i}^{j}$ for $j=1,\dots,d$, if the summary statistic is continuous or approximately continuous (e.g.\ large counts), a KDE of the $j$-th marginal density is given by

\begin{equation} \label{eq:KDE}
\hat{g}_{j} (x) = \dfrac{1}{n} \sum_{i=1}^{n} K_{h} (x - s_{i}^{j}),
\end{equation}

where $K_{h}(\cdot)$ is a non-negative kernel function with bandwidth parameter $h$. See \citet{Izenman1991} for a review and discussion about KDE. Similarly, the corresponding estimated cumulative distribution function $\hat{G}_{j}(\cdot)$ can be obtained. The choice of the kernel function could be various. \citet{An2018} select the Gaussian kernel for simplicity and for its unbounded support.

By using KDE, the semiBSL estimator can accommodate non-normal marginal summary statistics. For the second aspect, semiBSL uses a Gaussian copula to model dependence between summaries. The density function for the Gaussian copula is given by

\begin{equation*}
c_{\vect{R}} (\vect{u}) = \dfrac{1}{|\vect{R}|} \exp \Big\{ -\dfrac{1}{2} \vect{\eta}^{\top} (\vect{R}^{-1} - \vect{I}_{d}) \vect{\eta} \Big\}, 
\end{equation*}

where $\vect{R}$ is the correlation matrix, $\vect{\eta} = (\Phi^{-1}(u_{1}),\dots,\Phi^{-1}(u_{d}))^{\top}$, $\Phi^{-1}(\cdot)$ is the inverse CDF of a standard normal distribution, and $\vect{I}_{d}$ is a $d$-dimensional identity matrix. The main appeal of the Gaussian copula here is that it is fast to estimate the correlation matrix (see details later). Combining the advantages from both aspects, we are then able to have a tractable likelihood estimator written as $p_{ssl} (\vect{s}_{y} | \vect{\phi}_{ssl})$, where $\vect{\phi}_{ssl} = (\vect{R},u_{1},\dots,u_{d},g_{1} (s_{y}^{1}),\dots,g_{d} (s_{y}^{d}))$ is estimated by $\hat{\vect{\phi}}_{ssl} = (\hat{\vect{R}},\hat{u}_{1},\dots,\hat{u}_{d},\hat{g}_{1} (s_{y}^{1}),\dots,\hat{g}_{d} (s_{y}^{d}))$, where $\hat{u}_{j} = \hat{G}_{j} (s_{y}^{j})$ and $\hat{g}_{j} (s_{y}^{j})$ is given by equation \eqref{eq:KDE}. The full semiBSL density estimator is given by

\begin{equation} \label{eq:likelihood_semiBSL}
p_{ssl} (\vect{s}_{y} | \hat{\vect{\phi}}_{ssl}) = \dfrac{1}{\sqrt{\hat{\vect{R}}}} \exp \Big\{ -\dfrac{1}{2} \hat{\vect{\eta}}_{\vect{s}_{\vect{y}}}^{\top} (\hat{\vect{R}}^{-1} - \vect{I}_{d}) \hat{\vect{\eta}}_{\vect{s}_{\vect{y}}} \Big\} \prod_{j=1}^{d} \hat{g}_{j} (s_{y}^{j}).
\end{equation}

Finally we provide details on obtaining $\hat{\vect{R}}$. With a collection of simulations $\{\vect{\eta}_{\vect{s}_{i}} \}_{i=1}^{n}$, it can either be estimated with maximum likelihood estimation, or as \citet{An2018} advocate, one could use the Gaussian rank correlation (GRC, \citet{Boudt2012}). The GRC estimator is known to be more robust. The $(i,j)$-th entry of the GRC matrix is given by

\begin{equation*} \label{eq:grc}
\hat{\rho}_{i,j}^{\mathrm{grc}} = \dfrac{\sum_{k=1}^n \Phi^{-1}\Big(\dfrac{r(s_k^i)}{n+1}\Big) \Phi^{-1}\Big(\dfrac{r(s_k^j)}{n+1}\Big)}
{\sum_{k=1}^n \Phi^{-1} \Big( \dfrac{k}{n+1} \Big)^2},
\end{equation*}

where $r(\cdot): \mathbb{R} \rightarrow \mathcal{A}$, and $\mathcal{A} \equiv \{1,\ldots,n\}$, is the rank function. Note that here the GRC is used to estimated the correlation matrix in semiBSL. However, our package also permits the use of the GRC in standard BSL to provide more robustness. In doing so, a transformation to the covariance matrix is needed afterwards.

We point out that when the data generation process is complicated, most of the computational cost will be spent on model simulations. Thus the computational overhead introduced by the semi-parametric likelihood estimator may be negligible relative to the standard SL estimator.

\subsection{An Accelerated Likelihood Estimator with Shrinkage Estimation} \label{subsec:SL_shrinkage}

We can see from equations \eqref{eq:likelihood_BSL} and \eqref{eq:likelihood_semiBSL} that both estimators involve estimation of a covariance or correlation matrix. The number of free parameters in the covariance matrix grows quadratically with the number of summary statistics. Thus, for a high-dimensional summary statistic, a large number $n$ of simulations may be required to estimate the likelihood with reasonable precision.

Here we introduce two shrinkage estimators of the covariance matrix that have shown to be useful in empirical examinations \citep{An2019, Ong2018a, An2018}. The shrinkage estimators can be applied to BSL and semiBSL. Note that we skip uBSL because the unbiasedness property will be violated with shrinkage estimation. The selection of the penalty parameter will be discussed later in Section \ref{subsec:others}. For clarification, we change the notations in this subsection temporarily and restore them back after. Let $x_{1},\dots,x_{n}$ be samples from a multivariate normal distribution $\mathcal{N}(\vect{\mu},\vect{\Sigma})$. The dimension of each sample is $d$. $\vect{S}$ is the sample covariance and $\vect{\Theta} = \vect{\Sigma}^{-1}$ is the precision matrix.

\subsubsection{Graphical Lasso}

The graphical lasso \citep{Friedman2008} aims to find a sparse precision matrix by maximising the following $l_1$ regularised log-likelihood

\begin{equation*}
\log p(x) = c + \log|\vect{\Theta}| - tr(\vect{\Theta} \vect{S}) - \lambda||\vect{\Theta}||_{1},
\end{equation*}

where $c$ is a constant, $||\cdot||_{1}$ is the $l_1$ norm, and $\lambda$ is the penalty parameter. The problem is solved with convex optimisation.  The penalty parameter $\lambda$ controls the sparsity in $\vect{\Theta}$, where large $\lambda$ leads to high sparsity and vice versa. We refer the readers to \citet{Friedman2008} for more details about the graphical lasso. In our package we use the graphical lasso implementation in the \proglang{R} package \pkg{glasso} \citep{Rpkg:glasso}.

\subsubsection{Warton's Estimator}

\citet{Warton2008} introduces a straightforward shrinkage estimator that is extremely fast to compute but still performs well. Define $\hat{\vect{D}}$ as the diagonal matrix of $\vect{S}$, then the sample correlation matrix can be computed with $\hat{\vect{C}} = \hat{\vect{D}}^{-1/2} \vect{S} \hat{\vect{D}}^{-1/2}$. The Warton's shrinkage for the correlation matrix is given by

\begin{equation*}
\hat{\vect{C}}_{\gamma} = \gamma \hat{\vect{C}} + (1 - \gamma) \vect{I}_{d},
\end{equation*}

where $\gamma \in [0,1]$ is the shrinkage parameter, $\vect{I}_{d}$ is the identity matrix. The correlation matrix degenerates to the identity matrix as $\gamma$ approaches $0$. With a correlation to covariance conversion, we get the Warton's covariance estimator for $\vect{\Sigma}$

\begin{equation*}
\hat{\vect{\Sigma}}_{\gamma} = \hat{\vect{D}}^{1/2} \hat{\vect{C}}_{\gamma} \hat{\vect{D}}^{1/2}.
\end{equation*}

\subsection{Other Implementation Details} \label{subsec:others}

\subsubsection{Incorporating BSL with MCMC}

Once we have selected a likelihood function estimator, we can use a Bayesian sampling algorithm to sample from the BSL approximation to the posterior $p(\vect{\theta} | \vect{s}_{\vect{y}})$. This is currently achieved using MCMC in our package. Pseudo-code for the MCMC BSL algorithm is provided in Algorithm \ref{alg:MCMC_BSL}.

The number of simulations $n$ affects the variance of the likelihood estimator, which is very important in an MCMC algorithm. We want $n$ to be large enough so that the MCMC chain is very unlikely to get stuck at an overestimated likelihood value, but not so large that the computational efficiency is hindered. The value of $n$ can be loosely chosen depending on the number of free parameters in the likelihood function and adjusted based on the acceptance rate and MCMC effective sample size (ESS). Empirical evidence \citep{Price2018, An2018} has shown that the approximate posterior distribution (BSL, uBSL and semiBSL) is remarkably insensitive to $n$, provided that $n$ is not too small in which case the Monte Carlo error of the MCMC approximation can be large. \citet{Price2018} demonstrate empirically that aiming for a standard deviation of the synthetic log-likelihood estimator of around $1.5-2$ at a representative parameter value provides the most computational efficient results for standard BSL.

We recommended that the initial value of the Markov chain have non-negligible support under the BSL posterior.  Our experience with MCMC BSL is that if the chain is initialised far in the tails of the BSL posterior, then MCMC BSL exhibits slow convergence.  The initial value for the Markov chain may be sourced from experts, previous analyses in the literature or a short run of another likelihood-free algorithm.

\begin{algorithm}[!htp] \label{alg:MCMC_BSL}
\SetKwInOut{Input}{Input}
\SetKwInOut{Output}{Output}
\Input{Summary statistic of the observed data $\vect{s_y}$; the prior distribution $p(\vect{\theta})$, the proposal distribution $q(\cdot | \cdot)$; the number of simulations used to estimate synthetic likelihood $n$; the number of iterations $M$; and the initial value of the chain $\vect{\theta}^{0}$.}
\Output{MCMC sample $(\vect{\theta}^{0},\vect{\theta}^{1}, \ldots, \vect{\theta}^{M})$ from the BSL posterior, $p_{A}(\vect{\theta} | \vect{s_{y}})$.  Some samples can be discarded as burn-in if required.}

Generate $n$ datasets with the simulation function $\{\vect{x}_{i}\}_{i=1}^{n} \overset{iid}{\sim} p(\cdot | \vect{\theta}^{0})$\\
Compute the corresponding summary statistics $\{\vect{s}_{i}\}_{i=1}^{n} = \{\vect{S} (\vect{x}_{i})\}_{i=1}^{n}$\\
Estimate the auxiliary parameter $\hat{\vect{\phi}}$ depending on the SL estimator\\
Compute the estimated SL $\hat{p}_{A}^{0} = \hat{p}_{A} (\vect{s}_{\vect{y}} | \hat{\vect{\phi}})$ with equation \eqref{eq:likelihood_BSL}, \eqref{eq:likelihood_uBSL} or \eqref{eq:likelihood_semiBSL}\\
\For{$i = 1$ to $M$}{
Propose a parameter value with $\vect{\theta}^{*} \sim q(\cdot|\vect{\theta}^{i-1})$\\
Generate $n$ datasets with the simulation function $\{\vect{x}_{i}^{*}\}_{i=1}^{n} \overset{iid}{\sim} p(\cdot | \vect{\theta}^{*})$\\
Compute the corresponding summary statistics $\{\vect{s}_{i}^{*}\}_{i=1}^{n} = \{\vect{S} (\vect{x}_{i}^{*})\}_{i=1}^{n}$\\
Estimate the auxiliary parameter $\hat{\vect{\phi}}$ depending on the SL estimator\\
Compute the estimated SL $\hat{p}_{A}^{*} = \hat{p}_{A} (\vect{s}_{\vect{y}} | \hat{\vect{\phi}})$ with equation \eqref{eq:likelihood_BSL}, \eqref{eq:likelihood_uBSL} or \eqref{eq:likelihood_semiBSL}\\

Compute $r=\min\left(1,\dfrac{\hat{p}_{A}^{*} \, p(\vect{\theta}^{*}) \, q(\vect{\theta}^{i-1}|\vect{\theta}^{*})}   {\hat{p}_{A}^{i-1} \, p(\vect{\theta}^{i-1}) \, q(\vect{\theta}^{*}|\vect{\theta}^{i-1})}\right)$\\
\eIf{$\mathcal{U}(0,1)<r$}{
Set $\vect{\theta}^{i}=\vect{\theta}^{*}$ and $\hat{p}_{A}^{i} = \hat{p}_{A}^{*}$\\
}{
Set $\vect{\theta}^{i}=\vect{\theta}^{i-1}$ and $\hat{p}_{A}^{i} = \hat{p}_{A}^{i-1}$\\
}
\caption{MCMC BSL algorithm.}
\bigskip
}
\end{algorithm}

\subsubsection{Penalty Selection} \label{subsec:pen_selection}

\citet{An2019} introduce a penalty selection approach for BSLasso (BSL with graphical lasso). Our goal is to reduce the value of $n$ required while still maintaining a similar level of noise in the SL approximation. Denote $\sigma$ as the standard deviation of the penalised log-likelihood estimator that we are aiming for. The value of $\sigma$ is typically around $1.5$, as suggested earlier. The algorithm for selecting the penalty parameter is given in Algorithm \ref{alg:select_penalty}. Here we use $\lambda$ for the penalty parameter for both the graphical lasso and Warton's estimator for notational convenience.

\begin{algorithm}[htp] \label{alg:select_penalty}
\SetKwInOut{Input}{Input}
\SetKwInOut{Output}{Output}
\Input{The BSL method to be used, either BSL or semiBSL, denoted \code{method}; the shrinkage estimation to be used, either graphical lasso or Warton, denoted \code{shrinkage}; parameter value with non-negligible posterior support $\vect{\theta}_{0}$; the number of simulations $n$; a number of potential penalty values, $\lambda_1,\ldots,\lambda_K$; the standard deviation of the log-likelihood estimator to aim for $\sigma$; and the number of repeats $M$.}
\Output{The selected penalty parameter $\lambda$.}
\For{$m = 1$ to $M$}{
Generate $n$ datasets with the simulation function $\{\vect{x}_{i}\}_{i=1}^{n} \overset{iid}{\sim} p(\cdot | \vect{\theta}_{0})$\\
Compute the corresponding summary statistics $\{\vect{s}_{i}\}_{i=1}^{n} = \{\vect{S} (\vect{x}_{i})\}_{i=1}^{n}$\\

\For{$k=1$ to $K$} {
\If{\code{method} is BSL} {
Compute the penalised covariance matrix $\vect{\Sigma}_{n,\lambda_k}$ using the shrinkage method of \code{shrinkage} with the samples $\{\vect{s}_{i}\}_{i=1}^{n}$\\
Compute the sample mean $\vect{\mu}_{n} (\vect{\theta})$\\
Set $\hat{\vect{\phi}}_{sl} = (\vect{\mu}_{n} (\vect{\theta}), \vect{\Sigma}_{n,\lambda_k})$\\
Estimate the log SL $l_{m,k} = \log p_{sl} (\vect{s}_{\vect{y}} | \hat{\vect{\phi}}_{sl})$ with equation \eqref{eq:likelihood_BSL}\\
}
\If{\code{method} is semiBSL} {
Compute and penalise the Gaussian rank correlation matrix using the shrinkage method of \code{shrinkage} with the samples $\{\vect{s}_{i}\}_{i=1}^{n}$, and save it to $\vect{R}_{n,\lambda_k}$\\
Compute $\hat{u}_{1},\dots,\hat{u}_{d},\hat{g}_{1} (s_{y}^{1}),\dots,\hat{g}_{d} (s_{y}^{d})$\\
Set $\hat{\vect{\phi}}_{ssl} = (\vect{R}_{n,\lambda_k},\hat{u}_{1},\dots,\hat{u}_{d},\hat{g}_{1} (s_{y}^{1}),\dots,\hat{g}_{d} (s_{y}^{d}))$\\
Estimate the log SL $l_{m,k} = \log p_{ssl} (\vect{s}_{\vect{y}} | \hat{\vect{\phi}}_{ssl})$ with equation \eqref{eq:likelihood_semiBSL}\\
}
}
}
Compute $\sigma_{k}$ as the standard deviation of $\{l_{m,k}\}_{m=1}^{M}$ for $k=1,\dots,K$\\
Choose the $\lambda$ for which the empirical standard deviation $\sigma_{k}$ is closest to $\sigma$.
\bigskip
\caption{Procedure to select the penalty value $\lambda$ for BSL or semiBSL with shrinkage estimation.}
\end{algorithm}

\section{Using the BSL Package} \label{sec:package}

Two main functionalities that are offered in the \pkg{BSL} package are (1) running an MCMC algorithm with a chosen SL estimator to sample from the approximate posterior distribution, (2) selecting the penalty parameter with the given SL estimator and shrinkage method. Parallel computing is supported with the \proglang{R} package \pkg{foreach} \citep{Rpkg:foreach} so that the $n$ independent model simulations can be run on a multi-core computer. In this section, we use the built-in example \code{ma2} to show the usage of functions.

\subsection{Description of the MA(2) Example} \label{subsec:ma2}

%The moving-average (MA) time series model has been studied prevalently in ABC and SL literatures (REFS). 

Here we consider the MA(2), moving average of order $2$, time series model with parameter $\vect{\theta} = (\theta_{1}, \theta_{2})$. The parameter space is constrained to $\{\vect{\theta} : -1 < \theta_{2} < 1,\,\theta_{1} + \theta_{2} > -1,\,\theta_{1} - \theta_{2} < 1\}$ so that the process is stationary and invertible. The model has two favorable properties as an illustrative example, (1) the likelihood is tractable and hence sampling from the true posterior is feasible for comparisons, (2) the data generation process is fast. The model can be described with the following equation

\begin{equation*}
y_{t} = z_{t} + \theta_{1} z_{t-1} + \theta_{2} z_{t-2}, \text{ for $t = 1,\dots,T$,}
\end{equation*}

where $z_t \overset{iid}{\sim} \mathcal{N}(0,1)$ for $t=-1,0,\dots,T$ is white noise. The prior is set to be uniform on the feasible region of the parameter space, and zero elsewhere. We set $T = 50$ and use the raw data as the summary statistic. Note that the likelihood function is exactly multivariate normal with mean zero, $\mathrm{Var}(y_{t}) = 1 + \theta_{1}^{2} + \theta_{2}^{2}$, $\mathrm{Cov}(y_{t},y_{t-1}) = \theta_{1} + \theta_{1} \theta_{2}$, $\mathrm{Cov}(y_{t},y_{t-2}) = \theta_{2}$ and all other covariances are $0$. We load the observed data and simulation function with 

\begin{Schunk}
\begin{Sinput}
R> data("ma2", package = "BSL")
\end{Sinput}
\end{Schunk}

\subsection{The Model Object} \label{subsec:model}

We first introduce the S4 class \code{BSLModel} that encloses all ingredients for our model; namely a simulation function (\code{fnSim}), a summary statistic function (\code{fnSum}), a point estimate or initial value of the parameter (\code{theta0}) and other optional arguments for the model we want to run (for example \code{simArgs} and \code{fnLogPrior} as given below). The first three must be provided to generate a valid \code{BSLModel} object. This is done by running the creator function \code{BSLModel}. For example, we run the following command for the MA(2) model

\begin{Schunk}
\begin{Sinput}
R> model <- BSLModel(fnSim = ma2_sim, fnSum = function(x) x, 
+                    simArgs = list(T = 50), theta0 = c(0.6, 0.2), 
+                    fnLogPrior = ma2_logPrior)
\end{Sinput}
\begin{Soutput}
*** initialize "BSLMODEL" ***
has simulation function: TRUE 
has summary statistics function: TRUE 
has initial guess / point estimate of the parameter: TRUE 
running a short simulation test ... 
*** end initialize ***
\end{Soutput}
\end{Schunk}

We can see in the console message that upon initialisation, the \code{BSLModel} creator function automatically checks if the three requisites exist, and if one or more elements are missing the function will only create an empty object. By default the creator function also checks if the functions can run properly by running $10$ simulations as a short test. If the test takes too long or is undesired, it can be turned off with the flag \code{test = FALSE}. \code{theta0} is the initial guess of the parameter value, and will be used as the starting value in MCMC. Next we discuss how to properly code the simulation, summary statistic and prior function. The model object is mostly created for internal usage, so the related S4 methods are not described here.

\subsubsection{The Simulation Function}

A fast simulation function is helpful for the simulation-based methods in this paper. The simulation function is coded by the user and must have the model parameter $\vect{\theta}$ as its first argument. Other necessary arguments can be put after $\vect{\theta}$ if needed. The output can be of any valid \proglang{R} class and will be sent to the corresponding summary statistic function as the input data directly.

For example, the built-in simulation function of the MA(2) example is a valid simulation function. The length of the time series is determined by the second argument $T$, which can be set to a fixed value with \code{simArgs = list(T = 50)}. \code{simArgs} must be a list of the additional arguments to be put into the simulation function \code{fnsim}.

\begin{Schunk}
\begin{Sinput}
R> ma2_sim
\end{Sinput}
\begin{Soutput}
function (theta, T) 
{
    rand <- rnorm(T + 2)
    y <- rand[3:(T + 2)] + theta[1] * rand[2:(T + 1)] + theta[2] * 
        rand[1:T]
    return(y)
}
<bytecode: 0x0000000004864c90>
<environment: namespace:BSL>
\end{Soutput}
\end{Schunk}

For most complex likelihood-free applications, the simulation process will not be as straightforward as the MA(2) example. If the model simulation is complex and time-consuming, we encourage users to write their own simulation function in \proglang{C}/\proglang{C++} and use \pkg{Rcpp} \citep{Eddelbuettel2011} to call the function within \proglang{R}, as the running time of the algorithm is typically dominated by model simulations. For a reference text on writing functions in \pkg{Rcpp}, see \citet{Eddelbuettel2013}. The \code{ma2\_sim} function generates a single model realisation. If it is fast to produce more than one independent model simulation, for example if code vectorisation can be implemented, the \pkg{BSL} package also allows users to write the simulation function that directly simulates $n$ independent model simulations simultaneously. The code below shows how can this can be implemented for the MA(2) simulation function.

\begin{Schunk}
\begin{Sinput}
R> ma2_simVec = function(n, theta, T) {
+      rand <- matrix(rnorm(n * (T + 2)), n, T + 2)
+      y <- rand[, 3 : (T + 2)] + theta[1] * rand[, 2 : (T + 1)] 
+      + theta[2] * rand[, 1 : T]
+      sapply(1 : n, function(i) list(y[i, ]))
+  }
\end{Sinput}
\end{Schunk}

The vectorised simulation function can be loaded with argument \code{fnSimVec}; if it is loaded the vectorised simulation function will be used regardless of whether \code{fnSim} is provided or not. The input of \code{fnSimVec} must follow the order of $n$, $\vect{\theta}$ and additional arguments. Note that vectorised simulation is incompatible with parallel computing described later.

\subsubsection{The Summary Statistic Function}

For simplicity, we load the identity function that returns the same data as the summary statistic in the MA(2) example. Similar to \code{fnSim}, \code{fnSum} also takes additional arguments if desired. The additional arguments for the summary statistic function need to be put right after the data. These arguments can be saved as a list to \code{sumArgs}. The return argument of the summary statistic function must be a $d$-dimensional vector of summary statistics.

\subsubsection{The Log Prior Function}

\code{fnLogPrior} computes the log density of the prior distribution. If the prior function is not provided, an improper flat prior will be used by default. However, in practice, defining a proper prior distribution is always recommended. This function should only take $\vect{\theta}$ as its input argument, and the output must not be positive infinity. Note that the normalisation constant of the prior distribution can be ignored, as we do below for the MA(2) example.

\begin{Schunk}
\begin{Sinput}
R> ma2_logPrior
\end{Sinput}
\begin{Soutput}
function (theta) 
{
    log(theta[2] < 1 & sum(theta) > -1 & diff(theta) > -1)
}
<bytecode: 0x000000000485bd68>
<environment: namespace:BSL>
\end{Soutput}
\end{Schunk}

\subsection{The Main Function} \label{subsec:main_fn}

The primary function of the package is \code{bsl}, which uses MCMC to draw samples from an approximation to the idealised BSL posterior distribution $p_{A} (\vect{\theta} | \vect{s}_{\vect{y}})$. The type of likelihood estimator described in Section \ref{sec:methods} (BSL, uBSL and semiBSL) can be specified by the argument \code{method}. Shrinkage estimation on the covariance matrix in BSL and the correlation matrix in semiBSL can be specified with \code{shrinkage}. The minimal requirements of the \code{bsl} function include the observed data (\code{y}), the number of simulations for synthetic likelihood estimation (\code{n}), total number of MCMC iterations (\code{M}) and the covariance matrix of the normal random walk Metropolis algorithm (\code{covRandWalk}). For the MA(2) example, a minimalistic call to the main function that runs MCMC BSL (using the estimator of equation \eqref{eq:likelihood_BSL}) is given by the following command.

\begin{Schunk}
\begin{Sinput}
R> resultMa2BSL <- bsl(y = ma2$data, n = 500, M = 300000, model = model, 
+      covRandWalk = ma2$cov, method = 'BSL', verbose = TRUE)
\end{Sinput}
\end{Schunk}

Similarly, we can run MCMC uBSL and semiBSL by changing the \code{method} option, 

\begin{Schunk}
\begin{Sinput}
R> resultMa2uBSL <- bsl(y = ma2$data, n = 500, M = 300000, model = model, 
+      covRandWalk=ma2$cov, method = 'uBSL', verbose = TRUE)
\end{Sinput}
\end{Schunk}
\begin{Schunk}
\begin{Sinput}
R> resultMa2SemiBSL <- bsl(y = ma2$data, n = 500, M = 300000, model = model, 
+      covRandWalk=ma2$cov, method = 'semiBSL', verbose = TRUE)
\end{Sinput}
\end{Schunk}

The normal random walk covariance matrix, \code{covRandWalk}, should be defined based on the parameterisation used in the MCMC sampling; if a parameter transformation is employed, the covariance matrix should be specified accordingly.

\subsubsection{Shrinkage of the Likelihood Estimator}

Implementation of shrinkage methods described in Section \ref{subsec:SL_shrinkage} can be specified with arguments \code{shrinkage} (\code{\char`\"glasso\char`\"} for the graphical lasso estimator and \code{\char`\"Warton\char`\"} for Warton's estimator) and \code{penalty} for the penalty value ($\lambda$ in the graphical lasso and $\gamma$ in Warton's estimator). Note that the shrinkage takes place on the covariance matrix for method \code{BSL}, and on the correlation matrix (of the Gaussian copula) for method \code{semiBSL}. 

\begin{figure}[t!]
\centering
\includegraphics{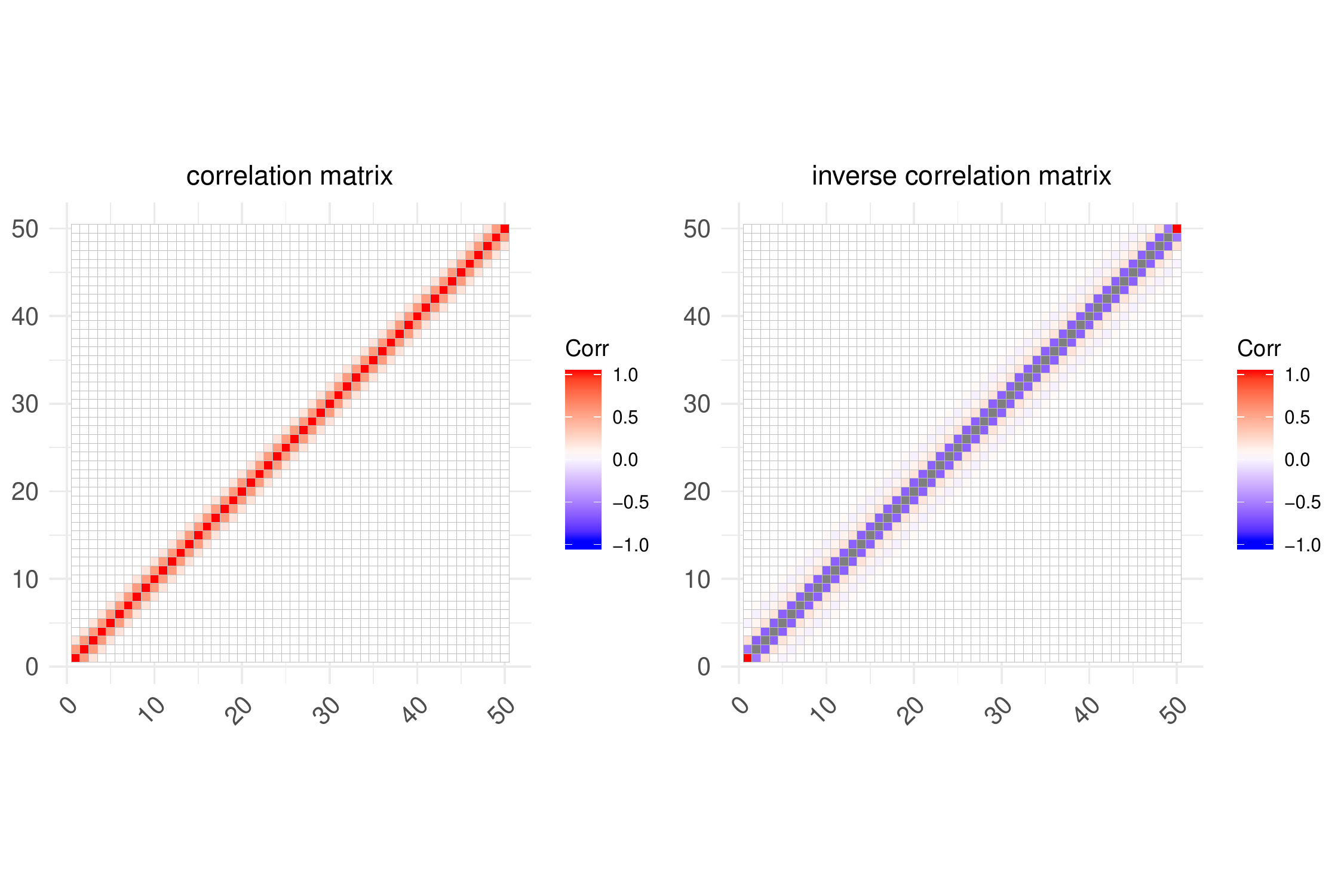}
\caption{\label{fig:cor_ma2} The correlation matrix (left) and inverse correlation matrix (right) of the true likelihood function for the MA(2) example.}
\end{figure}

Shrinkage should only be used when there is a relatively large number of entries near zero in the precision matrix for graphical lasso and covariance matrix for Warton's approach. This can be checked by inspecting the inverse correlation matrix or the correlation matrix at a representative parameter value. In the MA(2) example, we are able to calculate the true covariance matrix. For more complex models the covariance matrix can be estimated with a large number of model simulations at a reasonable parameter value. Here we use the \code{ggcorrplot} function in \pkg{ggcorrplot} \citep{Rpkg:ggcorrplot} to visualise the matrices. The visualisation of the correlation matrix and inverse correlation matrix for the MA(2) example are shown in Figure \ref{fig:cor_ma2}.

Both of the figures suggest that the correlation and inverse correlation matrices are sparse. It is also recommended to check the level of sparsity by computing the proportion of partial correlations below a certain threshold. For example, $81\%$ of the partial correlations between the summary statistics of the MA(2) example are below $0.01$. Thus, for this example, shrinkage estimation is expected to reduce the number of model simulations required for estimating the synthetic likelihood whilst not sacrificing much on posterior accuracy. We will describe how to select the penalty parameter in our \pkg{BSL} package in Section \ref{subsec:selectPenalty}. The following code runs MCMC BSL with the graphical lasso and Warton's shrinkage estimator.

\begin{Schunk}
\begin{Sinput}
R> resultMa2BSLasso <- bsl(y = ma2$data, n = 300, M = 300000, model = model, 
+      covRandWalk=ma2$cov, method = 'BSL', shrinkage = 'glasso', 
+      penalty = 0.027, verbose = TRUE)
R> resultMa2BSLWarton <- bsl(y = ma2$data, n = 300, M = 300000, model = model, 
+      covRandWalk=ma2$cov, method = 'BSL', shrinkage = 'Warton', 
+      penalty = 0.75, verbose = TRUE)
\end{Sinput}
\end{Schunk}

\subsubsection{Parameter Transformation}

If the prior distribution is bounded or the normal random walk cannot explore the parameter space efficiently, parameter transformation is recommended. Recall $p(\vect{\theta})$ is the prior distribution for $\vect{\theta} \in \vect{\Theta} \subseteq \mathbb{R}^{p}$. Suppose the parameters are independent and bounded, the prior function can be decomposed as

\begin{equation*}
p(\vect{\theta}) = \prod_{i=1}^{p} p_{i} (\theta_{i}), \text{ for $\theta_{i} \in (a_{i}, b_{i})$}, 
\end{equation*}

where $a_{i}$ and $b_{i}$ are the lower and upper bound for $\theta_{i}$. A straightforward but fruitful $1-1$ transformation that maps the range of the parameter to the real line is the logit transformation, which is given by

\begin{equation*}
\tilde{\theta_{i}} = \log \dfrac{\theta_{i} - a_{i}}{b_{i} - \theta_{i}} \text{, for $i = 1, \dots, p$}.
\end{equation*}

If the parameter is only bounded on one side, the transformation degenerates to a log transformation, i.e.\ $\tilde{\theta_{i}} = \log \dfrac{\theta_{i}}{b_{i} - \theta_{i}}$ if $a_{i}$ is $-\infty$ and $\tilde{\theta_{i}} = \log \dfrac{\theta_{i} - a_{i}}{\theta_{i}}$ if $b_{i}$ is $\infty$. The \code{bsl} function provides an easy-to-use log/logit transformation for independent and bounded parameters. The argument \code{logitTransformBound} is a $p$ by $2$ matrix containing the lower and upper bounds for each parameter. The only argument that needs to be changed is \code{covRandWalk}. There is no need to do any reparameterisation of the prior or simulation functions.

It is also possible to code a customised parameter transformation by editing the simulation function directly. If so the prior function should also be changed subject to the reparameterisation.

\subsubsection{Parallel Computation}

As a simulation-based method, the computational cost of BSL is mostly driven by the speed of the simulation process. Parallel computation is vital in complex applications where vectorisation is not possible. In our \pkg{BSL} package, the $n$ independent model simulations can be distributed across workers on a multi-core machine. The \proglang{R} package \pkg{doParallel} \citep{Rpkg:doParallel} provides a way to set up the cpu cores for parallel computation. For example, the following code determines all the available cores of a cpu and sets up the clusters for parallel jobs.

\begin{Schunk}
\begin{Sinput}
R> ncores <- detectCores()
R> cl <- makeCluster(detectCores())
R> registerDoParallel(cl)
\end{Sinput}
\end{Schunk}

This should be turned on prior to running the main \code{bsl} function if parallel computing is desired. To enable parallel computing with in \pkg{BSL}, set \code{parallel} to \code{TRUE} in the \code{bsl} function. We utilise the function \code{foreach} of package \pkg{foreach} \citep{Rpkg:foreach} to run parallel simulations in our \pkg{BSL} package. All other arguments supported by \code{foreach} can be passed with \code{parallelArgs} in the \code{bsl} and \code{selectPenalty} functions. For illustration, the following code runs MCMC BSL with parallel computing,

\begin{Schunk}
\begin{Sinput}
R> resultMa2BSLParallel <- bsl(y = ma2$data, n = 500, M = 300000, 
+      model = model, covRandWalk = ma2$cov, method = 'BSL', 
+      parallel = TRUE, verbose = TRUE)
\end{Sinput}
\end{Schunk}

Note that parallel computing introduces additional communication time between workers. If the model simulation process is straightforward, parallel computing might increase the overall running time rather than reduce it. Thus this is not recommended in the MA(2) example. Vectorisation can be more effective in such cases. Once parallel computing is completed, the following code shuts down the parallel cores.

\begin{Schunk}
\begin{Sinput}
R> stopCluster(cl)
R> registerDoSEQ()
\end{Sinput}
\end{Schunk}

\subsection{Interpret and Visualise the BSL Result}

The output of the function \code{bsl} is saved as an S4 object \code{BSL}, which includes \code{theta} (approximate posterior samples), \code{loglike} (the MCMC chain of estimated log-likelihood values), \code{acceptanceRate} (acceptance rate of MCMC) for inspection of the Markov chain, as well as several other arguments which help to analyse the result. A full list of the returned values can be checked with \code{help(bsl, package = `BSL')}. We provide the basic \code{show}, \code{summary} and \code{plot} methods for the \code{BSL} class.

\begin{figure}[h!]
\centering
\begin{Schunk}
\begin{Sinput}
R> plot.ts(resultMa2BSL@loglike)
\end{Sinput}
\end{Schunk}
\includegraphics{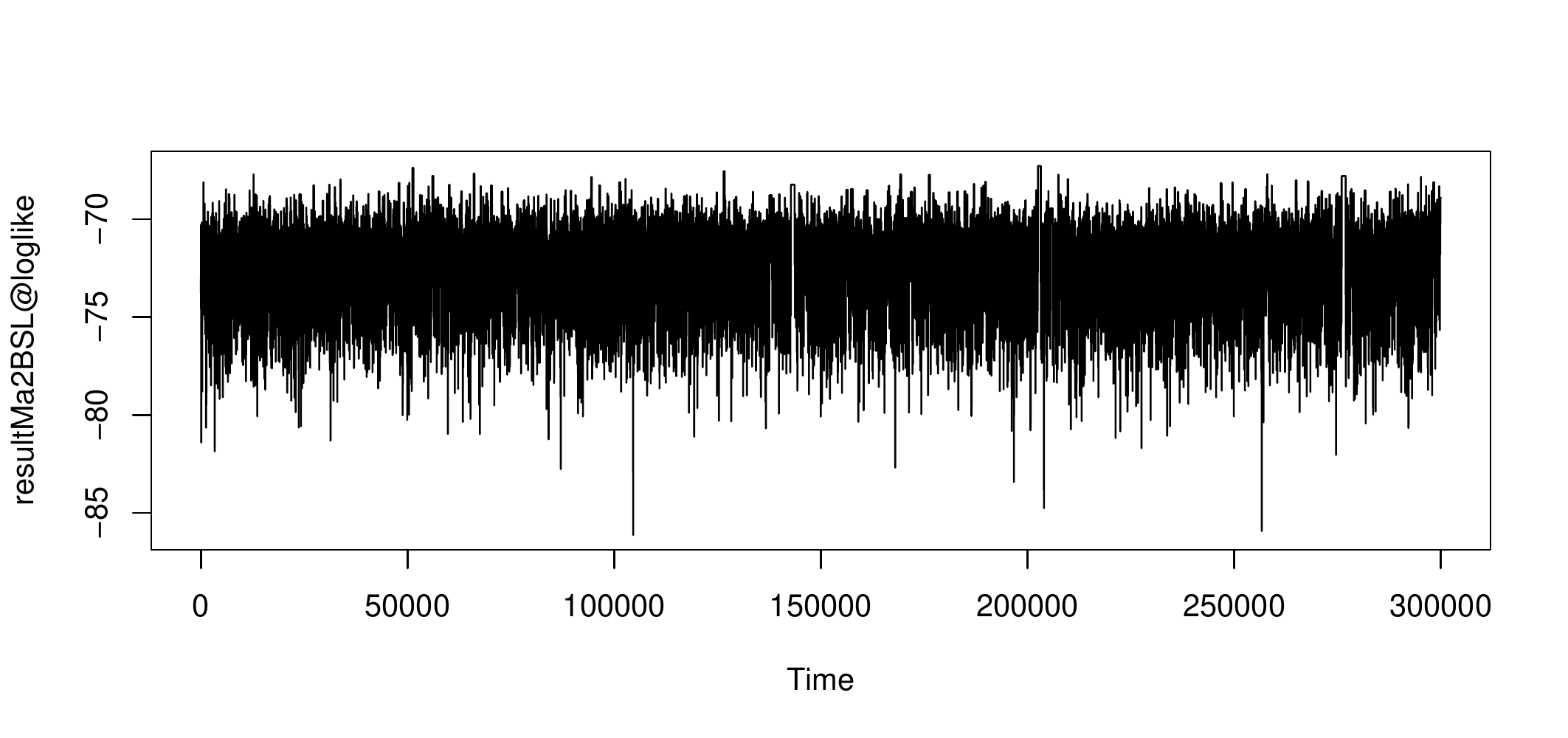}
\caption{\label{fig:ma2_traceplot} Trace plot of the synthetic log-likelihood estimates for the MA(2) example.}
\end{figure}

The following code provides some common MCMC diagnostics for the MCMC BSL result of the MA(2) example. The column title ESS in the \code{summary} result stands for the effective sample size of the approximate posterior samples. Here we only show the trace plot of the synthetic log-likelihood estimates (Figure \ref{fig:ma2_traceplot}), however we also recommend other specialised \proglang{R} packages (such as \pkg{coda} \citep{Rpkg:coda} and \pkg{plotMCMC} \citep{Rpkg:plotMCMC}) for other visualisations and quantitative diagonostics of MCMC convergence. 

\begin{Schunk}
\begin{Sinput}
R> show(resultMa2BSL)
\end{Sinput}
\begin{Soutput}
Call:
bsl(y = ma2$data, n = 500, M = 3e+05, model = model, covRandWalk = ma2$cov, 
    method = "BSL", verbose = TRUE)

Summary of theta:
      theta[1]             theta[2]           
[1,]  Min.   :-0.1001      Min.   :-0.549391  
[2,]  1st Qu.: 0.4648      1st Qu.: 0.006416  
[3,]  Median : 0.5749      Median : 0.109592  
[4,]  Mean   : 0.5732      Mean   : 0.109172  
[5,]  3rd Qu.: 0.6782      3rd Qu.: 0.217373  
[6,]  Max.   : 1.1953      Max.   : 0.719013  
Summary of loglikelihood:
   Min.  1st Qu.   Median     Mean  3rd Qu.     Max.  
 -86.13   -72.82   -71.54   -71.66   -70.35   -67.29  
Acceptance Rate:
[1]  0.1444
Early Rejection Rate:
[1]  0.001103
\end{Soutput}
\begin{Sinput}
R> summary(resultMa2BSL)
\end{Sinput}
\begin{Soutput}
              n acc. rate (\\%)    ESS theta[1]    ESS theta[2] 
            500              14            5530            5128 
\end{Soutput}
\end{Schunk}

\begin{figure}[h!]
\centering
\begin{Schunk}
\begin{Sinput}
R> plot(resultMa2BSL, which = 1, thetaTrue = c(0.6, 0.2), thin = 30)
R> mtext('Approximate Univariate Posteriors', line = 1, cex = 1.5)
\end{Sinput}
\end{Schunk}
\includegraphics{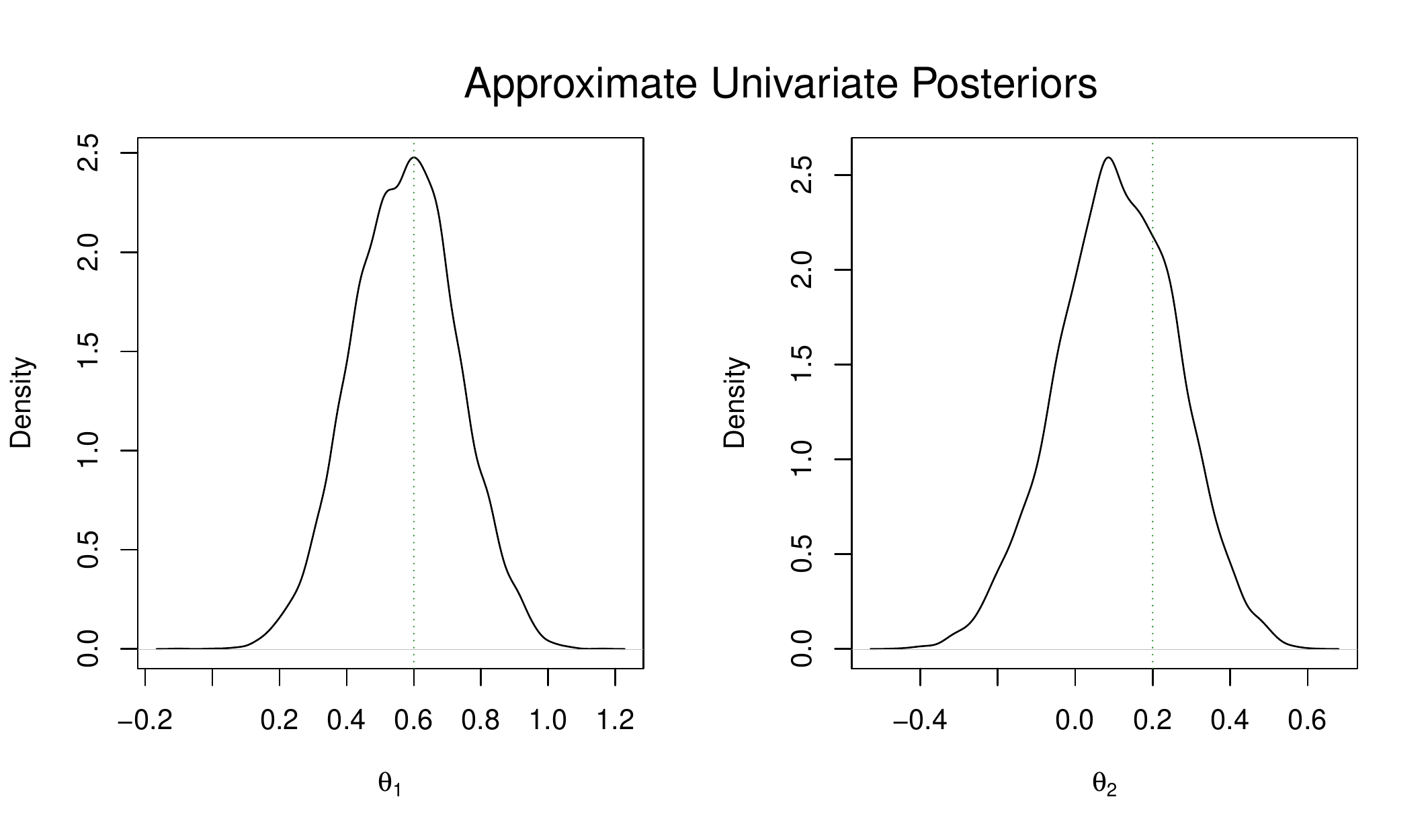}
\caption{\label{fig:ma2_posterior1} Approximate posterior distributions using the BSL estimator for the MA(2) example with \pkg{graphics}.}
\end{figure}

\begin{figure}[h!]
\centering
\begin{Schunk}
\begin{Sinput}
R> plot(resultMa2BSL, which = 2, thetaTrue = c(0.6, 0.2), thin = 30, 
+      options.density = list(color = 'coral4', fill = 'coral', alpha = 0.5),
+      options.theme = list(panel.background = element_rect(fill = 'beige'), 
+      plot.margin = grid::unit(rep(0.05,4), "npc")))
\end{Sinput}
\end{Schunk}
\includegraphics{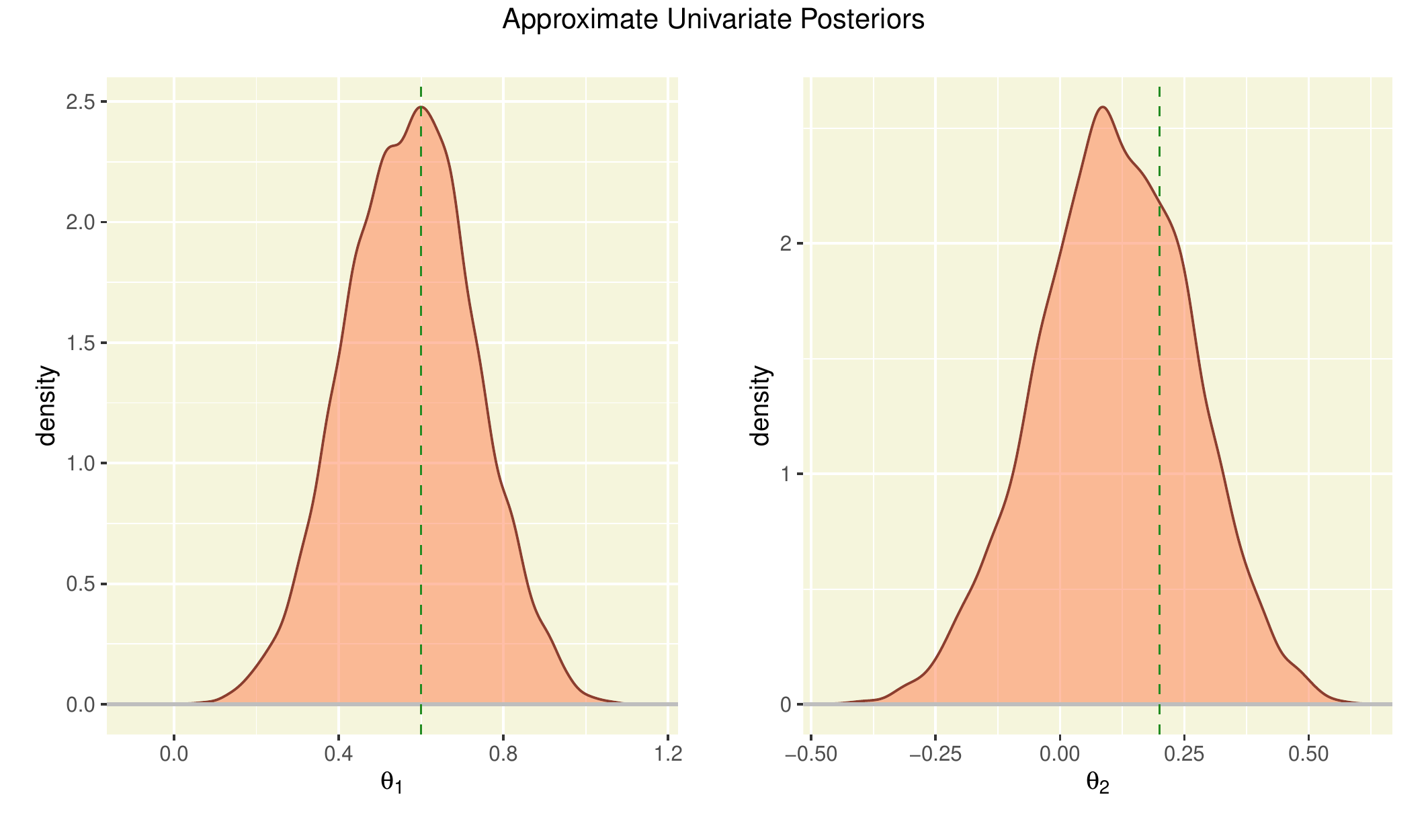}
\caption{\label{fig:ma2_posterior2} Approximate posterior distributions using the BSL estimator for the MA(2) example with \pkg{ggplot2}.}
\end{figure}

The \code{plot} function for \code{BSL} draws the approximate univariate posterior distribution for each parameter with either the default \proglang{R} \code{graphics} (Figure \ref{fig:ma2_posterior1}) or \code{ggplot2} (Figure \ref{fig:ma2_posterior2}).

It is often of interest to compare multiple \code{bsl} results at the same time, for example, the following code summarises the BSL, uBSL, semiBSL, BSL with graphical lasso and BSL with Warton results.

\begin{Schunk}
\begin{Sinput}
R> results <- list(resultMa2BSL, resultMa2uBSL, resultMa2SemiBSL, 
+                  resultMa2BSLasso, resultMa2BSLWarton)
R> labels <- c('bsl', 'uBSL', 'semiBSL', 'bslasso', 'bslWarton')
R> names(results) <- labels
R> t(sapply(results, summary))
\end{Sinput}
\begin{Soutput}
            n acc. rate (\\%) ESS theta[1] ESS theta[2]
bsl       500              14         5530         5128
uBSL      500              14         5023         5368
semiBSL   500              13         4874         5075
bslasso   300              28         8436         8038
bslWarton 300              31         9423         7319
\end{Soutput}
\end{Schunk}

It can be seen that BSL, uBSL and semiBSL have very similar acceptance rates and ESS values, which indicates that their performance is very close. Both shrinkage methods increase the acceptance rate and ESS with smaller $n$ compared to BSL.

\begin{figure}[h!]
\centering
\begin{Schunk}
\begin{Sinput}
R> par(mar = c(5, 4, 1, 2), oma = c(0, 1, 2, 0))
R> combinePlotsBSL(results, which = 1, thetaTrue = c(0.6, 0.2), thin = 30,
+      col = c('black', 'red', 'blue', 'green', 'cyan'), lty = 1:5, lwd = 1)
R> mtext('Approximate Univariate Posteriors', outer = TRUE, cex = 1.5)
\end{Sinput}
\end{Schunk}
\includegraphics{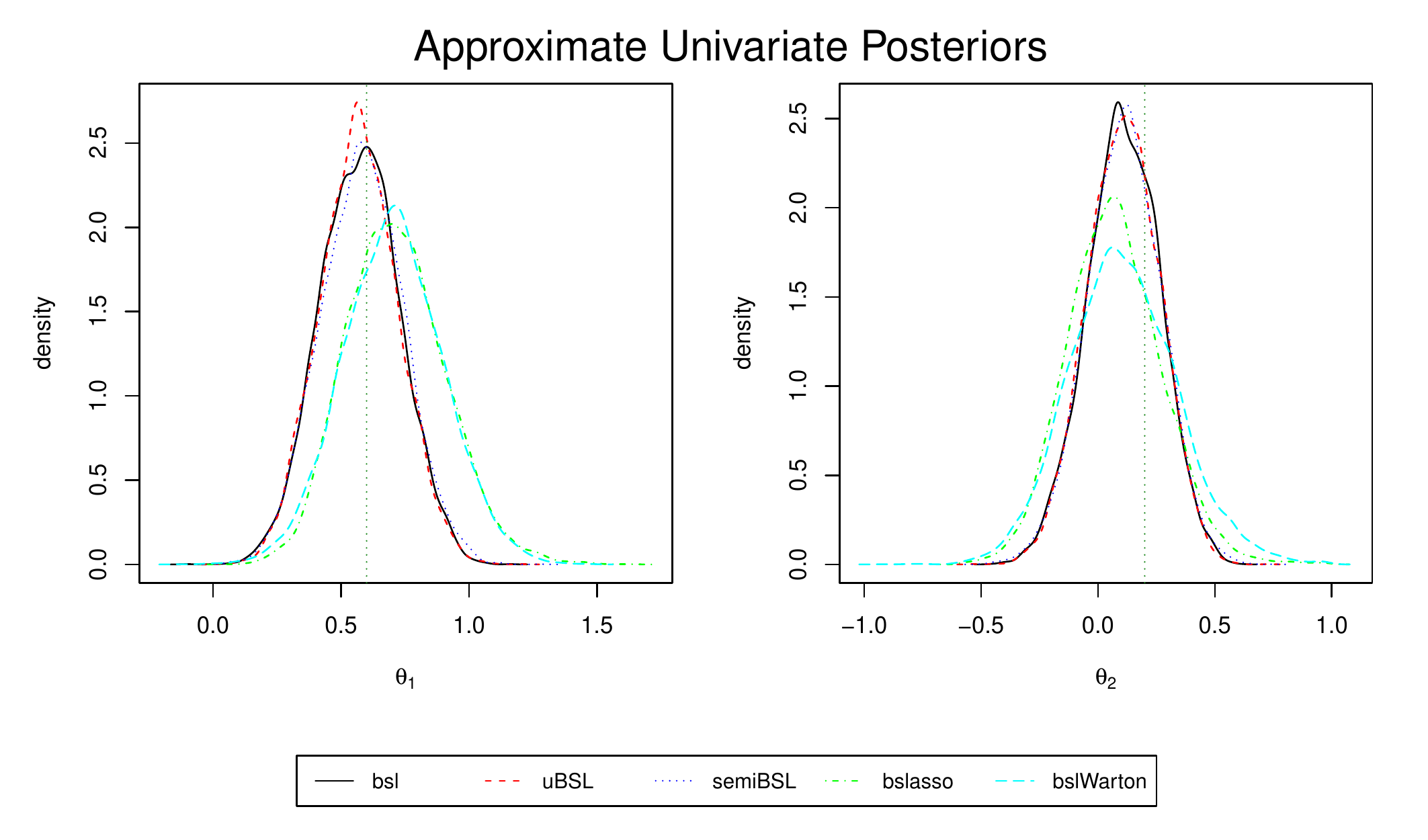}
\caption{\label{fig:ma2_posteriors1} Approximate posterior distributions using various BSL estimators for the MA(2) example with \pkg{graphics}.}
\end{figure}

\begin{figure}[h!]
\centering
\begin{Schunk}
\begin{Sinput}
R> combinePlotsBSL(results, which = 2, thetaTrue = c(0.6, 0.2), thin = 30,
+      options.color = list(values=c('black', 'red', 'blue', 'darkgreen', 'cyan')),
+      options.linetype = list(values = 1:5), 
+      options.size = list(values = rep(1, 5)),
+      options.theme = list(plot.margin = grid::unit(rep(0.03,4), "npc"),
+          axis.title = ggplot2::element_text(size=12), 
+          axis.text = ggplot2::element_text(size = 8),
+          legend.text = ggplot2::element_text(size = 12)))
\end{Sinput}
\end{Schunk}
\includegraphics{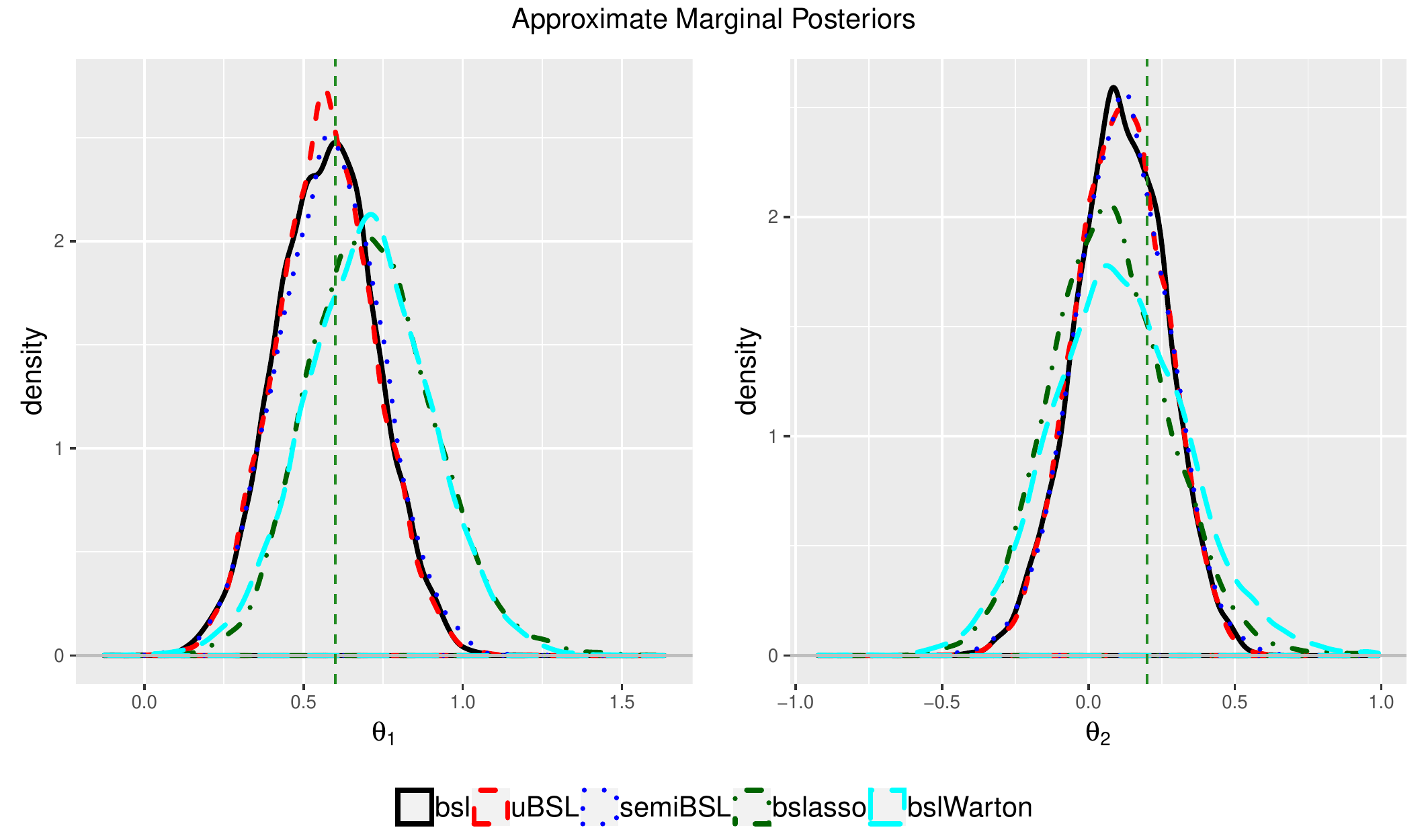}
\caption{\label{fig:ma2_posteriors2} Approximate posterior distributions using various BSL estimators for the MA(2) example with \pkg{ggplot2}.}
\end{figure}

The function \code{combinePlotsBSL} plots multiple approximate posterior distributions in the same figure. The arguments are similar to the \code{plot} function of \code{BSL}, except the object to be plotted should be a list of multiple \code{BSL} objects. Figure \ref{fig:ma2_posteriors1} and \ref{fig:ma2_posteriors2} give an example of the \code{combinePlotsBSL} function.

\subsection{Selecting the Penalty Parameter for Shrinkage} \label{subsec:selectPenalty}

\begin{figure}[h!]
\centering
\begin{Schunk}
\begin{Sinput}
R> print(sp_bsl_glasso_ma2)
\end{Sinput}
\begin{Soutput}
Call:
selectPenalty(ssy = ssy, n = c(50, 150, 300, 500), lambda_all = lambda_all, 
    theta = ma2$start, M = 100, sigma = 1.5, model = model, method = "BSL", 
    shrinkage = "glasso")

Penalty selected based on the standard deviation of the loglikelihood:
     n penalty sigma
11  50 0.31415  1.44
29 150 0.07995  1.46
50 300 0.02718  1.50
68 500 0.00575  1.49
\end{Soutput}
\begin{Sinput}
R> plot(sp_bsl_glasso_ma2)
\end{Sinput}
\end{Schunk}
\includegraphics{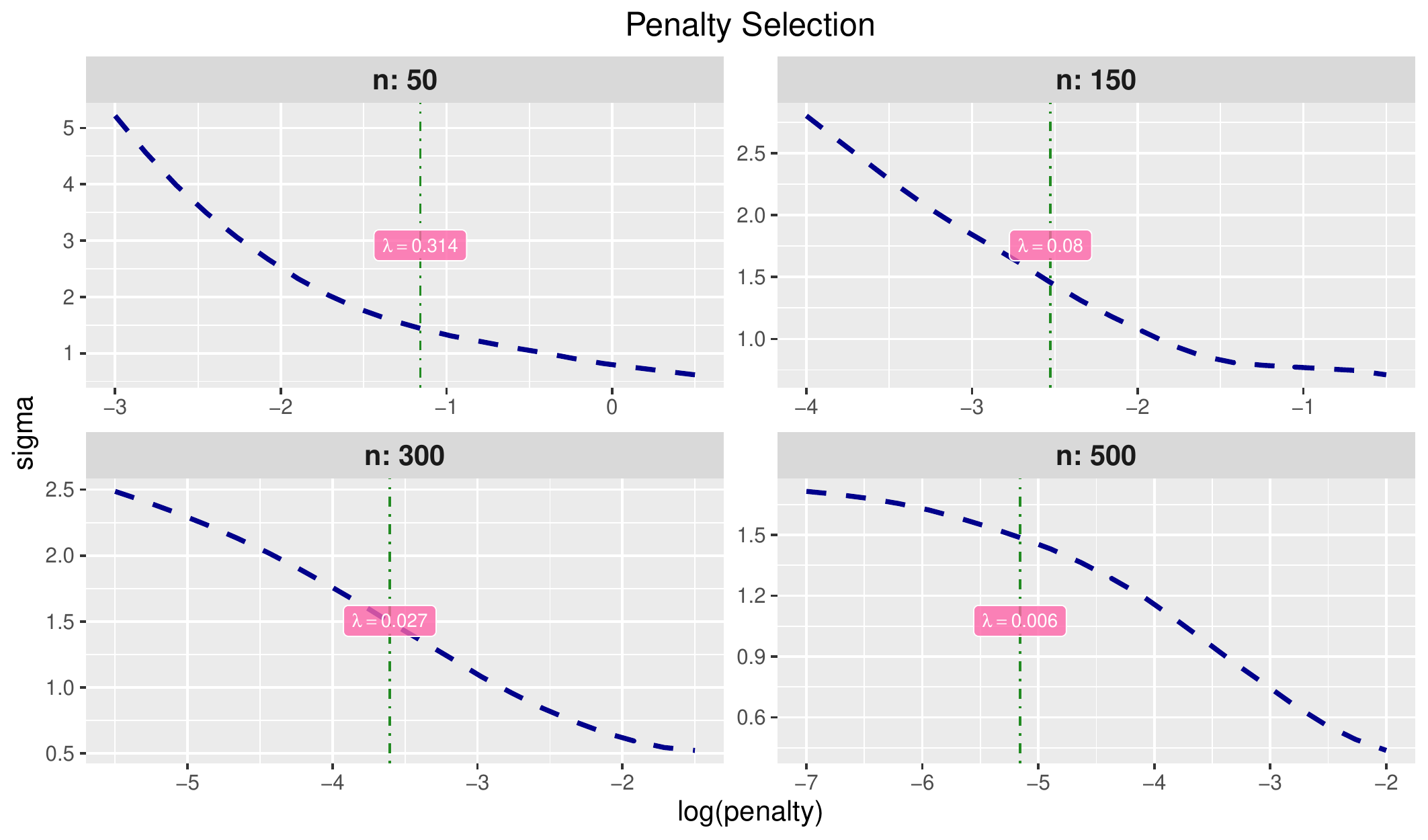}
\caption{\label{fig:ma2_select_pen_glasso} Penalty selecting result for BSL with graphical lasso for the MA(2) example.}
\end{figure}

If shrinkage is desired in estimating the SL, the corresponding penalty parameter value must be selected prior to running the \code{bsl} function. This can be done via the \code{selectPenalty} function. Multiple choices for the number of simulations $n$ can be tested at the same time; simulations from the largest value of $n$ are re-used for smaller values of $n$ by subsetting. The basic arguments of \code{selectPenalty} include the summary statistic of the observed data \code{ssy}; a vector of the number of simulations $n$ to test; a list of the candidate penalty values \code{lambda\_all} corresponding to each $n$; a point estimate of the parameter \code{theta}; the number of repeats \code{M}; the target standard deviation \code{sigma}; the model of interest \code{model}; the SL estimator \code{method} and the shrinkage estimation method \code{shrinkage}. Example code is given below (Figure \ref{fig:ma2_select_pen_glasso}) for selecting the $\lambda$ of glasso with the standard BSL estimator for the MA(2) example.

\begin{Schunk}
\begin{Sinput}
R> ssy <- ma2_sum(ma2$data)
R> lambda_all <- list(exp(seq(-3, 0.5, length.out = 20)), 
+                     exp(seq(-4, -0.5, length.out = 20)), 
+                     exp(seq(-5.5, -1.5, length.out = 20)), 
+                     exp(seq(-7, -2, length.out = 20)))
R> set.seed(100)
R> sp_bsl_glasso_ma2 <- selectPenalty(ssy = ssy, n = c(50, 150, 300, 500), 
+      lambda_all, theta = c(0.6, 0.2), M = 100, sigma = 1.5, model = model, 
+      method = 'BSL', shrinkage = 'glasso')
\end{Sinput}
\end{Schunk}

The result is of S3 class \code{penbsl}. This class has support for \code{print} and \code{plot} functions.

\section{Summary and Future Work} \label{sec:summary}

This paper has presented the first comprehensive software package for Bayesian synthetic likelihood methods.  It implements three different types of synthetic likelihood estimators (standard, unbiased and semi-parametric) and includes functionality for two different types of shrinkage covariance estimators to reduce the number of required model simulations.  The package includes functions for extracting useful statistics and visualisations of the results after running \code{bsl}.  It also includes several built-in examples to illustrate the functionality of the package.

Apart from the MA(2) example described in Section \ref{sec:package}, the package also include two other examples: a discrete-time stochastic cell biology model, \code{cell}; and a financial model for currency exchange data, \code{mgnk}. The former example features a high dimensional summary statistic, while the latter example has a relatively large number of parameters. Additional descriptions and example code can be found in the package help link. We plan to add additional applications in a future release, including real data.

In a future release we plan to incorporate the methods of \citet{Frazier2019}, which allows BSL to be much more computationally efficient when the model is misspecified, in particular when the model is unable to recover the observed value of the summary statistic.  All of the methods in our \pkg{BSL} package use MCMC to explore the parameter space, and can thus be slow when model simulation is computationally intensive and there is a large number of parameters.  Future research for this package could consider incorporating the variational Bayes synthetic likelihood methods \citep{Tran2017, Ong2018b}, which scale to a large number of summary statistics and/or parameters at the expense of assuming a multivariate normal approximation of the posterior.

We welcome \proglang{R} developers from worldwide to help contribute to the \pkg{BSL} package. We have made the source code available at \url{https://github.com/cran/BSL} to allow other researchers to contribute to the BSL package.

\section*{Computational details}

The results in this paper were obtained using
\proglang{R}~3.5.3 with the
\pkg{BSL}~3.0.0 package. \proglang{R} itself
and all packages used are available from the Comprehensive
\proglang{R} Archive Network (CRAN) at
\url{https://CRAN.R-project.org/}.

\section*{Acknowledgments}

CD was supported by an Australian Research Council's Discovery Early Career Researcher Award funding scheme (DE160100741).  ZA was supported by a scholarship under CDs grant DE160100741 and a top-up scholarship from the Australian Research Council Centre of Excellence for Mathematical and Statistics Frontiers (ACEMS). LFS would like to acknowledge support from QUT, ACEMS and EPSRC grant EP/S00159X/1. Computational resources and services used in this work were provided by the HPC and Research Support Group, Queensland University of Technology, Brisbane, Australia.

\bibliographystyle{apalike}
\bibliography{BSL}

\end{document}